\documentclass[useAMS,usenatbib]{mn2e}
\usepackage{epsfig,slashbox,pict2e}

\title[The PACO faint sample]{The {\it Planck}-ATCA Coeval Observations (PACO) project: the faint sample}
\author[L. Bonavera et al.]{
\parbox[t]{\textwidth}
{Laura Bonavera$^{1,2}$\thanks{E-mail: bonavera@sissa.it}, Marcella Massardi$^{3}$, Anna Bonaldi$^{3,4}$,\\ Joaquin Gonz{\'a}lez-Nuevo$^{1}$, Gianfranco De Zotti$^{3,1}$, Ronald D. Ekers$^{2}$}\\
\vspace*{8pt} \\
$^{1}$SISSA, via Bonomea 265, I--34136 Trieste, Italy\\
$^{2}$Australia Telescope National Facility, CSIRO Astronomy and Space Science, PO Box 76, Epping, NSW 1710, Australia\\
$^{3}$INAF-Ossevatorio Astronomico di Padova, vicolo dell'Osservatorio 5, I--35122 Padova, Italy\\
$^{4}$Jodrell Bank Centre for Astrophysics, School of Physics \& Astronomy, University of Manchester, Oxford Road, Manchester M13 9PL}

\begin{document}

\date{}
\pagerange{\pageref{firstpage}--\pageref{lastpage}} \pubyear{2010}
\maketitle
\label{firstpage}

\begin{abstract}
The {\it Planck}-ATCA Co-eval Observations (PACO) project collected data between 4.5 and 40 GHz for 482 sources selected within the Australia Telescope 20 GHz (AT20G) catalogue and observed with the Australia Telescope Compact Array (ATCA). Observations were done  almost simultaneously with the {\it Planck} satellite, in the period between July 2009 and August 2010. In this paper we present and discuss the data for the complete sample of 159 sources with $S_{\rm AT20G}>200$ mJy in the Southern ecliptic pole region.

The Planck Early Release Compact Source Catalogue (ERCSC) contains 57 of our sources. A comparison between the PACO catalogue and the ERCSC confirms that the reliability of the latter is better than 95 per cent. The missing ERCSC sources are typically associated with the Large Magellanic Cloud, the Milky Way or are otherwise extended. 

The spectral analysis of the PACO faint catalogue shows a spectral steepening of the sources at high frequencies, confirming the results obtained from the PACO bright sample.

A comparison with AT20G measurements, carried out, on average, a few years earlier, has demonstrated that, on these timescales, our sources show a rather high variability with an rms amplitude of $\simeq 40$ per cent at 20 GHz. The source spectral properties are found not to vary substantially with flux density, except for an increase of the fraction of steep spectrum sources at fainter flux densities.

Our data also allow us to extend by a factor $\simeq 5$ downwards in flux density the source counts at $\simeq 33\,$GHz and $\simeq 40\,$GHz obtained from {\it Planck}'s Early Release Compact Source catalogue. This allows us to substantially improve our control on the contribution of unresolved extragalactic sources to the power spectrum of small scale fluctuations in CMB maps.

\end{abstract}

\section{Introduction} \label{sec:intro}
The knowledge of the extragalactic radio sources at frequencies higher than 10 GHz is still poor, although the situation has been improving rapidly in recent years \citep[see][for a review]{2010A&ARv..18....1D}, thanks especially to the 9C \citep{2003MNRAS.342..915W,2010MNRAS.404.1005W}, AT20G \citep{2010MNRAS.402.2403M,MAS10}, 10C \citep{2010arXiv1012.3659D}, WMAP \citep{2009ApJS..180..283W,LO07,LO09,MAS09}, and {\it Planck} \citep{GEN} surveys. The study of extragalactic sources at millimeter wavelengths is important both per s\'e and in connection with CMB experiments, since these sources are the main astrophysical contaminant of CMB maps on angular scales smaller than $20'$--$30'$ at frequencies of up to $\simeq 100\,$GHz \citep{TOF98,TOF99,GF99,2005A&A...431..893D}.

{\it Planck} offers a unique opportunity to carry out an unbiased investigation of the spectral properties of radio sources in a poorly explored frequency range, partially unaccessible from the ground. A definition of sources Spectral Energy Distributions (SEDs) over a frequency range as large as possible is crucial in determining their physical properties and in identifying the different components that may contribute to their emission. Given that observations in the full {\it Planck} frequency range will not be repeatable at least in the foreseeable future, it is essential not to lose the occasion, while {\it Planck} is flying, of greatly increasing its science yields by simultaneous (i.e. not affected by variability) ground based observations at lower frequencies as well as at frequencies overlapping with {\it Planck} channels.

This has motivated the {\it Planck}-ATCA Co-eval Observations (PACO) project, that consisted in observations with the Australia Telescope Compact Array (ATCA) at several frequencies between 4.5 and 40 GHz, almost simultaneously with the {\it Planck} satellite.  In the period between July 2009 and August 2010 we observed 482 extragalactic sources, extracted from the Australia Telescope 20 GHz (AT20G) survey catalog \citep{2010MNRAS.402.2403M}. Of these, 344 sources form 3 partially overlapping complete sub-samples, selected for different purposes.
\begin{itemize}
\item The PACO faint sample, presented in this paper, is made of 159 sources with $S_{20{\rm GHz}}\ge 200$ mJy in the South Ecliptic Pole region (ecliptic latitude $<-75^\circ$) and with $3\hbox{h}<\hbox{RA}<9\hbox{h}$, $\delta<-30^\circ$. Near the Ecliptic Poles {\it Planck}'s scan circles intersect. Therefore the area is covered many times, and {\it Planck}'s sensitivity is maximal.

\item The PACO spectrally-selected sample comprises the 69 sources with $S_{20\rm{GHz}}>200$ mJy and inverted or upturning spectra in the frequency range 5--20 GHz, selected over the whole Southern sky;

\item The PACO bright sample comprising the 189 sources with $S_{20\rm{GHz}}>500\,$mJy at $\delta<-30^\circ$ \citep{MAS11}.
\end{itemize}
A full description of the PACO project and of its main goals is given in \cite{MAS11}. The aims specific to the PACO faint sample are:
\begin{itemize}
\item Extend to fainter flux densities the characterization of radio source spectra from 4.5 GHz to the {\it Planck} frequency range. Although {\it Planck}'s sensitivity is maximal in the selected area, we expect that only a minority of sources in this sample will be detected by {\it Planck}. The estimated completeness limits of the {\it Planck} Early Release Point Source Catalog \citep[ERCSC,][]{ERCSC} are of 1 and 1.5 Jy at 33 and 40 GHz, respectively \citep{statprop}, and the minimum flux density of an ERCSC extragalactic source  is of 480 mJy at 30 GHz and of 585 mJy at 44 GHz \citep{ERCSC}. Only 57 PACO faint sources have ERCSC flux density measurements at least at one frequency.  The final catalog is expected to reach deeper flux density limits \citep{LEA08} but, nevertheless, we will probably get the full {\it Planck} spectral coverage only for a subset of sources in this sample. However further spectral information can be obtained via a stacking analysis.

\item Extend the determination of source counts at $\simeq 33$ and $\simeq 40$ GHz obtained from the analysis of the ERCSC \citep{statprop} downwards in flux density by a factor of $\simeq 5$. Going down in flux density is important to control the contamination of CMB maps by faint radio sources.
\end{itemize}
The plan of the paper is as follows. In \S\,\ref{sec:sam} we summarize the main steps of data acquisition, reduction and calibration. In \S\,\ref{sec:sample} we describe the PACO faint sample. In \S\,\ref{sec:analysis} we analyze the spectral behaviour between 4.5 and 40 GHz of our sources, their variability, and estimate the differential source counts at 33 and 40 GHz. Our main results are summarized in \S\,\ref{sec:conclusion}.

\section{Observations and data reduction} \label{sec:sam}
A detailed description of observations, data reduction and calibration can be found in \cite{MAS11}. Here we present only a brief summary of the main points. The observations exploited the capabilities of the new Compact Array Broadband Backend (CABB, Ferris \& Wilson 2002, Wilson et al. 2011) system, that allows to observe $2\times2$ GHz simultaneous bands in continuum. Each band is split into $2048\times 1\,$MHz frequency channels. Applied to the $6\times 22\,$m antennas of the Australia Telescope Compact Array, the CABB gives a noise level per band reaching 0.5 mJy in 1 min on source down to the 7 mm wavelength band. We chose to use the receivers at 7mm (intermediate frequencies, IFs, at 33 and 39 GHz), to overlap the lower-frequency {\it Planck} channels, at 12mm (IFs at 18 and 24 GHz) to overlap our selection frequency, and at 3-6 cm (IFs at 5.5 and 9 GHz) to extend the SEDs to lower frequencies.

Since most of the sources are point-like we can observe with any array configuration. The dates of the observing runs, the telescope configurations and their properties are summarized in Tables 1 and 2 of \cite{MAS11}. The project was allocated 62 observing runs, for a total of $\sim450$ hours. For each run and for each frequency, we observed a bandpass calibrator and a primary calibrator. Thanks to the compactness of most of the PACO sources neither imaging nor phase calibration is necessary, but since sources are self-calibrated we get good images when observing with hybrid arrays. We got a suitable flux density estimation from the visibilities using the triple-correlation techniques. Each target source has been observed in a single 1.5 min-long shot for each frequency, that corresponds to a theoretical noise of less than 1 mJy.

Observations were generally made within 10 days from the {\it Planck} scan on any given source. An effort was made to observe sources in all the three bands (7mm, 12mm and 3-6cm) during the same run. Exceptions are due to bad weather conditions. The sources have been re-observed in several runs overlapping with the {\it Planck} observations (the satellite surveys the whole sky every 6 months), with a higher observing rate for the sources that appeared more variable.

For data reduction we used the MIRIAD software \citep{1995ASPC...77..433S}, recently updated to deal with CABB data. We inscribed MIRIAD tasks within a scripting pipeline for consistency in the analysis, and have independently reduced the data from each observing run and each frequency. After flagging the time intervals affected by bad weather or instrumental failures we have split the $\sim$2 GHz bands in 4$\times 512$ MHz sub-bands to exploit the spectral information. We have calibrated each sub-band for bandpass, flux density and leakages.

The classification of sources as point-like or extended was mostly taken from the AT20G catalogue and integrated at 7 mm by the analysis of the source phase closure and ratio between the data collected with long baselines and those collected with the whole array. The flux densities, and the associated errors, for point-like sources were computed using the triple product \citep{2001isra.book.....T}. For extended sources we used the scalar flux density on the shortest baseline which, however, is only a lower limit to the real flux density. We have in total 9 extended sources in the PACO faint sample, only 4 of which were classified as extended in the AT20G catalog.

At the end of the process, a quality check on the data has been performed comparing the data points with a polynomial fit to the source SEDs. This approach assumes that the SEDs of our sources are smooth across our frequency range. Misbehaving points are attributed to frequency dependent problems, not addressed by the automatic flagging that operates within each bands.

\section{The sample} \label{sec:sample}
\begin{figure}
\begin{center}
\includegraphics[width=0.5\textwidth]{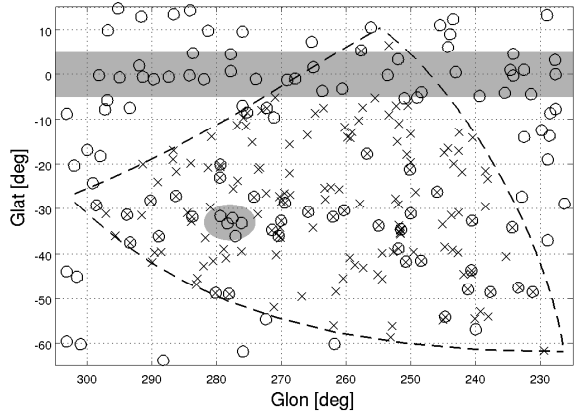}
\caption{The area covered by the PACO faint sample is that bounded by the dashed lines, excluding the grey areas, i.e. the Galactic plane region ($\vert b \vert \leq 5^\circ$) and the Large Magellanic Cloud (LMC) region (a circle of $4^\circ$ radius). The crosses show the positions of PACO sources, the circles those of 30 GHz ERCSC sources. }
  \label{fig:ERCSC}
\end{center}
\end{figure}

The region covered by the PACO faint sample is shown in Fig.~\ref{fig:ERCSC}. The crosses are the PACO faint sample sources, the circles are the 30 GHz ERCSC sources. There are 16 ERCSC sources within the region bounded by the dashed lines, not included in the present sample. Twelve lie either within the excluded Large Magellanic Cloud (PLCKERC030 G276.20-33.22, PLCKERC030 G277.16-36.06, PLCKERC030 G277.62-32.14, PLCKERC030 G278.40-33.32, PLCKERC030 G279.49-31.67) or $\vert b \vert < 5^\circ$ (PLCKERC030 G252.05+03.28, PLCKERC030 G260.67-03.20, PLCKERC030 G263.73-03.72, PLCKERC030 G265.14+01.48, PLCKERC030 G267.93-01.02, PLCKERC030 G253.63-00.23, PLCKERC030 G248.44-04.00) areas. Three (PLCKERC030 G271.21-09.56, PLCKERC030 G249.07-05.17, PLCKERC030 G251.04-05.33) are at low Galactic latitude and were not detected by the AT20G survey, suggesting that they are extended Galactic sources or peaks of the Galactic diffuse emission. The last one (PLCKERC030 G240.02-56.83) is Fornax A, missed by the AT20G because its core emission at 20 GHz is below the detection limit and its extended emission has a size exceeding the ATCA field-of-view \citep{MAS08}. Thus the fraction of potentially spurious ERCSC sources in the PACO faint area is $\le 5\%$.

\begin{table}
\center
\caption{Number of PACO faint sources present in the ERCSC, for each frequency channel.The last column gives the rms distances between ERCSC and PACO positions for the channels with a number of sources greater than 20.}
\label{tab:crossmatch}
\begin{tabular}{cccc}
\hline
freq & search radius & number of & rms distance \\
(GHz) & (arcmin) & sources & (arcmin)\\
\hline
30 & 15 & 39 & 2.6\\
44 & 12 & 27 & 2.1\\
70 & 7 & 29 & 1.3\\
100 & 5 & 42 & 1\\
143 & 4 & 44 & 0.8\\
217 & 3 & 31 & 0.3\\
353 & 3 & 15 & -\\
545 & 3 & 5 & -\\
857 & 3 & 3 & -\\
\hline
\end{tabular}
\end{table}

\begin{figure}
\begin{center}
\includegraphics[width=0.35\textwidth, angle=90]{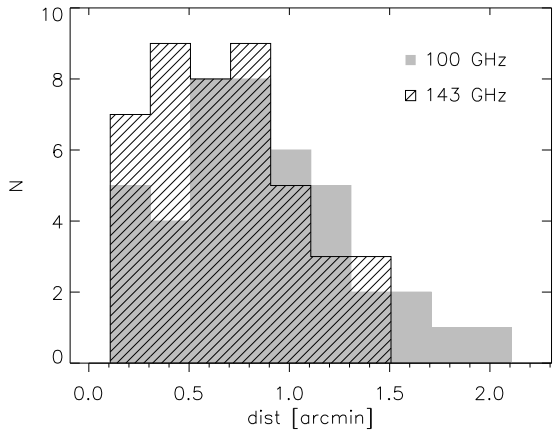}
\caption{Distribution of the angular distances (in arcmin) between the PACO and ERCSC positions of the sources resulting in the cross-match with the 100 (grey area) and 143 GHz (hatched area) {\it Planck} channels.}
  \label{fig:dist}
\end{center}
\end{figure}

\begin{figure*}
\begin{center}
\includegraphics[width=0.7\textwidth, angle=90]{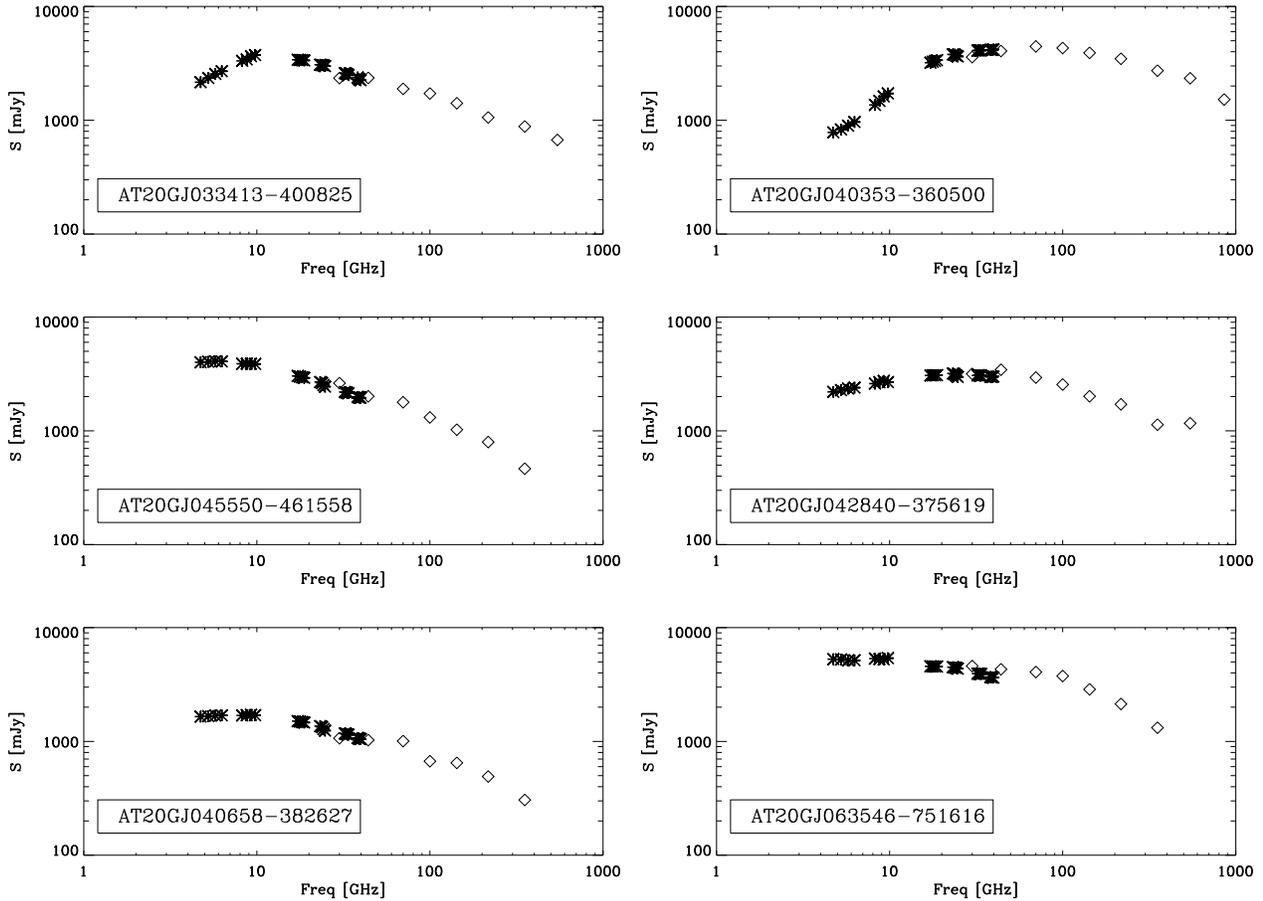}
\caption{Examples of radio spectra obtained combining PACO (asterisks) and ERCSC (diamonds) measurements.}
  \label{fig:allsed}
\end{center}
\end{figure*}

The results of the cross-match between PACO and ERCSC sources in all {\it Planck} channels are summarized in Tab.~\ref{tab:crossmatch}, where we also list the adopted search radii and the rms values of the distances between {\it Planck} and PACO positions in each channels. The rms values are much lower than the search radii, and consistent with the rms scatter in positions being lower than FWHM/5 as found in \cite{ERCSC}. Fig.~\ref{fig:dist} shows the distribution of the angular distances between the positions of the sources obtained by cross-matching the ERCSC catalogue with the PACO faint sample one: we chose the 100 GHz (grey area) and 143 GHz (hatched area) {\it Planck} channels, where we found the greatest number of ERCSC-PACO common objects. Of the 57 sources present in the ERCSC, 25 have measurements at more than 5 different frequencies. While a joint analysis of PACO and ERCSC results is deferred to a subsequent paper, we show in Fig.~\ref{fig:allsed} some examples of spectra determined combining the two data sets.

All the 159 sources in the PACO faint sample have been observed at least once at all the frequencies mentioned in \S\,\ref{sec:sam}. Bad weather conditions affect all the runs for 5 sources (J034205-370320, J040848-750720, J052006-454745, J062620-534151, J064118-355433) and it was not possible to repeat the observations. All of them are extended. Moreover, the results of mapping and mosaicing in total intensity and polarisation the core and the western lobe of Pictor A, namely J051926-454554 and J051949-454643, will be presented in a subsequent paper; these two sources will not be considered further here. Therefore the PACO faint catalogue, which is available as online supplementary material, contains 152 sources. The content of its columns is as follows.
\begin{description}
\item[1.] AT20G name;
\item[2.--3.] J2000 equatorial coordinates;
\item[4.] Epoch of observation;
\item[5.] Flag `s' for simultaneity with {\it Planck};
\item[6.] Flag `e' for extended sources;
\item[7.--31.] Flux densities in mJy for the 24 PACO sub-bands;
\item[32.--55.] Flux density errors;
\item[56.--60.] Double power-law fit parameters (see \S\,\ref{sec:analysis}).
\end{description}

\section{Data analysis} \label{sec:analysis}
The analysis was restricted to the 143 point-like sources. The PACO data (4.5-40 GHz) have been fitted with the double power-law
\begin{equation}
S(\nu)=S_0/[(\nu/\nu_0)^{-a}+(\nu/\nu_0)^{-b}]\label{dpl},
\end{equation}
with $\nu$ and $\nu_0$ in GHz, $S$ and $S_0$ in Jy. $S_0$, $\nu_0$, $a$ and $b$ are free parameters. For spectral indices we adopt the convention:
\begin{equation}
\alpha^{\nu_1}_{\nu_2}=\frac{\log(S_1/S_2)}{\log(\nu_1/\nu_2)}.
\end{equation}
The fit was performed for sources having more than 4 data points for each receiver band. This was the case for all but 2 sources. The fit was considered successful if the reduced $\chi^2<2.9$ \citep[see][]{MAS11}. The fit was not successful for nine sources which show an irregular spectrum that could not be described by the fitting formula in eq.~(\ref{dpl}). In most cases the irregularities of the spectrum seem to be due to residual calibration problems. When observations at more than one epoch are available, we chose the one with the greatest number of data points.

\subsection{Spectral behaviour} \label{sec:spec}
\begin{figure}
\begin{center}
\includegraphics[width=0.4\textwidth, angle=90]{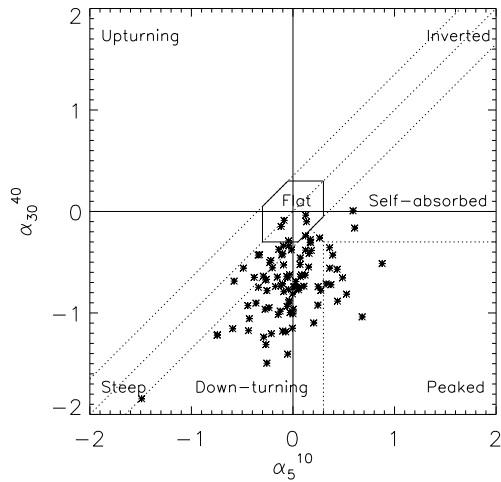}
\caption{Distribution of PACO faint sources with $200\,\hbox{mJy}\le S_{\rm AT20G} < 500\,$mJy in the $\alpha_5^{10}$--$\alpha_{30}^{40}$ plane. The dotted lines show the boundaries for the various spectral types.}
  \label{fig:color-color}
\end{center}
\end{figure}

The double-power-law fit that we have performed allows us to estimate the spectral indices for any frequency range between 4.5 and 40 GHz. Excluding sources brighter than 500\,mJy, which are included in the PACO bright sample, whose spectral properties are discussed by  \citet{MAS11}, we have 98 sources in the range $200\,\hbox{mJy}\le S_{\rm AT20G} < 500\,$mJy. The spectral classification, based on the spectral indices between 5 and 10 GHz, compared with those between 30 and 40 GHz ($\alpha_5^{10}$, $\alpha_{30}^{40}$), is illustrated in Fig.~\ref{fig:color-color}. The fractions of sources of each spectral type are listed in Table~\ref{tab:class1}, where the fraction found for the bright PACO sample are also given, for comparison. The main difference with the bright sample is the larger fraction of steep-spectrum sources, mostly at the expense of peaked- and flat-spectrum sources. 

\begin{table}
\caption{Percentages of sources of each spectral type in the PACO faint sample ($200 \le S_{\rm AT20G} < 500\,$ mJy) and in the bright sample ($S_{\rm AT20G} \ge500\,$mJy).}
\center
\label{tab:class1}
\begin{tabular}{|l|c|c|}
\hline
Type &$\!\!\!\!\!\!\!\!\!\!\!\!\!\!\!\! 200\hbox{mJy}\le S_{20\rm GHz}< 500\hbox{mJy}$  &  $S_{20\rm GHz}\ge  500\hbox{mJy}$ \\
  & per cent & per cent \\
\hline
flat          & $5.1 $    & $10.3 $ \\
steep         & $13.3 $   & $3.6 $ \\
inverted      & $0 $      & $0.6 $ \\
peaked        & $11.2 $   & $14.5 $  \\
down turning  & $65.3 $   & $66 $  \\
self absorbed & $5.1 $    & $4.8 $ \\
upturning     & $0 $      & $0 $   \\
\hline
\end{tabular}
\end{table}

For the 11 peaked sources with $200 \le S_{20\rm GHz}< 500$ mJy we computed the peak frequency as $\nu_p=\nu_0 (-b/a)^{(1/(b-a))}$, we found a median value of 16.6 GHz, a mean of 17.2 GHz and a dispersion of 3.7 GHz. For the 64 down-turning sources in the same flux density range we found a median break frequency value of 31.8 GHz, a mean of 26.2 GHz and a dispersion of 14.1. We note that 28 sources have a value of $\nu_0$ at the border of our frequency range and therefore not reliably estimated. This does not affect the value of the median.

\begin{figure}
\begin{center}
\includegraphics[width=0.35\textwidth, angle=90]{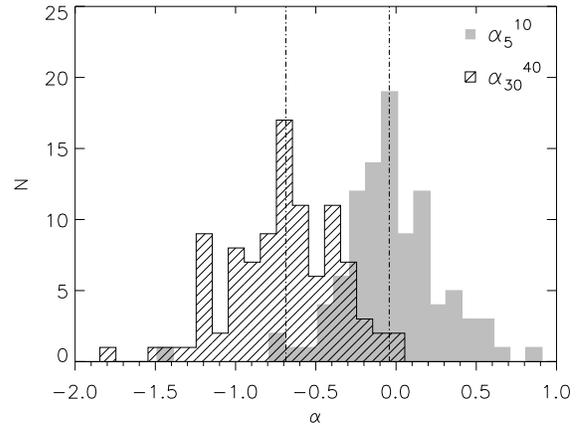}
\caption{Distributions of $\alpha_5^{10}$ (shaded area) and of $\alpha_{30}^{40}$ (hatched area). }
  \label{fig:histo}
\end{center}
\end{figure}

\begin{figure}
\begin{center}
\includegraphics[width=0.35\textwidth, angle=90]{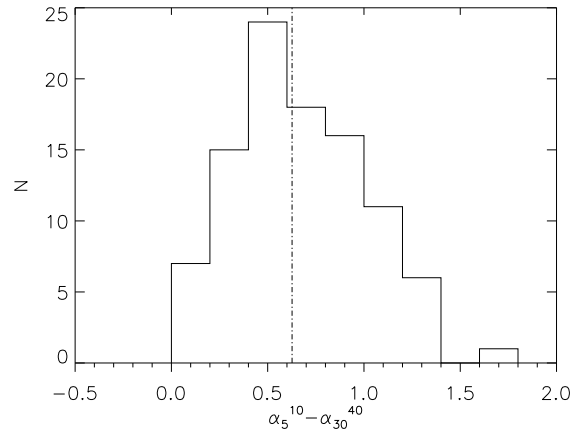}
\caption{Distribution of the difference $\alpha_5^{10} - \alpha_{30}^{40}$. }
  \label{fig:histo2}
\end{center}
\end{figure}

The distributions of spectral indices $\alpha_5^{10}$ and $\alpha_{30}^{40}$ are shown in Fig.~\ref{fig:histo}. The median values are $-0.04$ for $\alpha_5^{10}$ and $-0.69$ for $\alpha_{30}^{40}$, with interquartile ranges $[-0.22;0.15]$ and $[-0.93;-0.43]$ respectively, showing that the spectra are mainly flat at lower frequencies and  steepen at higher frequencies. The distribution of the differences between the two spectral indices shown in Fig.~\ref{fig:histo2} has mean value of $0.67\pm 0.04$ and standard deviation of 0.35; the median value is $0.63 \pm 0.04$. A steepening of the spectral indices between 20 and 90 GHz has also been reported by \cite{SAJ11}, based on VLA observations of a sample selected from the AT20G survey. Moreover a steepening of spectral indices at higher frequencies is also shown in the analysis of the NEWPS sample \citep{GO08} and the ERCSC sample \citep{statprop}.

We did the same analysis for the spectral indices between 5 and 20 GHz and between 20 and 40 GHz. The mean value of $\alpha_5^{20}$ is $-0.16$ with standard deviation of 0.31;  the mean value of $\alpha_{20}^{40}$ is $-0.60$ with a  standard deviation of 0.31. These values are similar, although not equal to those found for the bright sample ($-0.07$ and $-0.55$, respectively, with standard deviations of 0.32 and 0.34). This indicates that the overall spectral behaviour up to 40 GHz is not strongly dependent on flux density in the range covered by these samples, especially between 20 and 40 GHz.

\subsection{Variability}\label{sec:var}
\begin{figure}
\begin{center}
\includegraphics[width=0.35 \textwidth, angle=90]{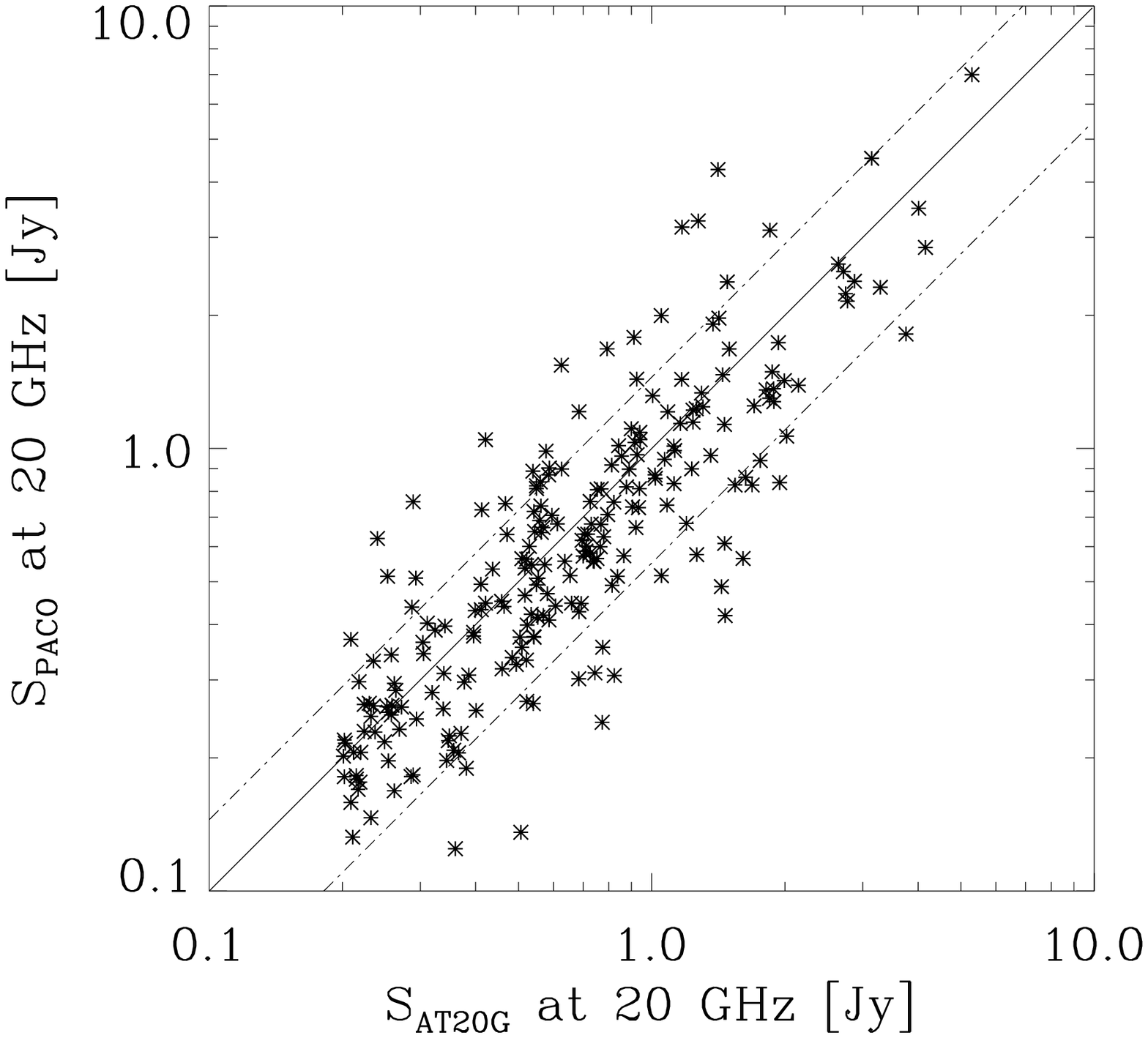}
\caption{Comparison between the AT20G flux densities at 20 GHz and the PACO ones scaled to 20 GHz. The solid line corresponds to $S_{\rm AT20G}=S_{\rm PACO}$ and the dot-dashed lines correspond to $S_{\rm PACO}=S_{\rm AT20G, 20GHz}(1 \pm \sigma_v)$.}
  \label{fig:cfr}
\end{center}
\end{figure}

The AT20G flux densities have been collected in several epochs between 2004 and 2008. Therefore, the time lag between the AT20G and the PACO observations ranges from several months to several years. Since variable sources are more likely to meet the detection limit of a blind survey if they are in a bright phase, it is not surprising that, on average, AT20G flux densities are slightly higher than the PACO ones. After correcting for the small frequency difference between the PACO and AT20G observations, using our measured spectral indices between 18.768 23.232 GHz, we find $\langle \log_{10}(S_{\rm AT20G}/S_{\rm PACO})\rangle =  0.043$, where $S_{\rm AT20G}$ and $S_{\rm PACO}$ are the AT20G flux density at 20 GHz and the PACO flux density scaled to 20 GHz. A comparison of the two flux densities is shown in Fig.~\ref{fig:cfr}. The least-squares linear relationship is $S_{\rm PACO}[Jy]= (0.89 \pm 0.04) \cdot S_{\rm AT20G}[Jy]+(0.05 \pm 0.05)$, showing no statistically significant deviations from a constant ratio between the two flux densities. It is important to note, in this respect, that the observations have been performed with the same front-end system, but the back-end has changed. Last, but not least, the new CABB system requires a more careful account of the spectral behaviour of the sources during the analysis, because of the broader bands. Our findings add to a series of indications of the high performance of the new system and show that there are no obvious instrumental effects that may complicate the comparison with measurements obtained before its upgrade.

\begin{figure}
\begin{center}
\includegraphics[width=0.35 \textwidth, angle=90]{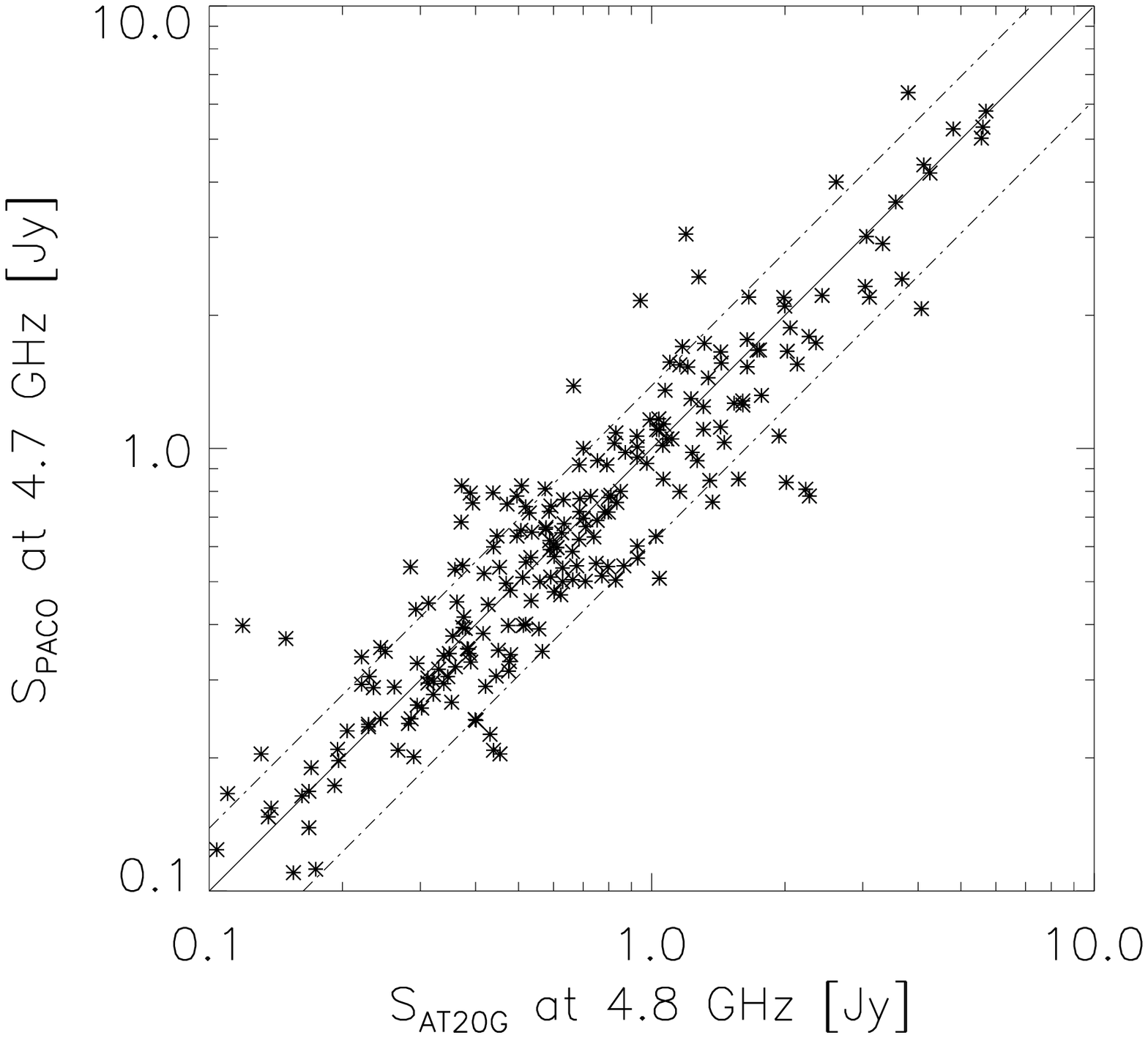}
\caption{Comparison between the AT20G flux densities at 4.8 GHz and the PACO ones at 4.744 GHz. The solid line corresponds to $S_{\rm AT20G}=S_{\rm PACO}$ and the dot-dashed lines correspond to $S_{\rm PACO, 4.7GHz}$=$S_{\rm AT20G, 4.8GHz}(1 \pm \sigma_v)$.}
  \label{fig:cfr_5}
\end{center}
\end{figure}

\begin{figure}
\begin{center}
\includegraphics[width=0.35 \textwidth, angle=90]{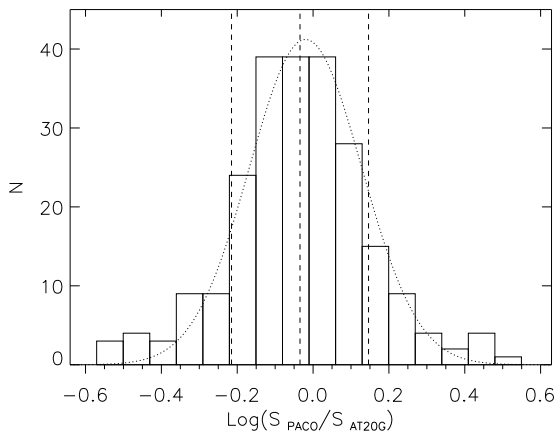}
\caption{Distribution of $\log_{10}(S_{\rm PACO}/S_{\rm AT20G})$ at $\simeq 20\,$GHz. The dotted line is the Gaussian best-fit. The dashed lines are the mean and the mean $\pm \sigma$ of the distribution.}
  \label{fig:gau}
\end{center}
\end{figure}

Since both the PACO and the AT20G measurement errors are negligibly small, the variability at $\simeq 20\,$GHz can be estimated as the standard deviation of the distribution of $S_{\rm PACO}/S_{\rm AT20G}$ or of $\log_{10}(S_{\rm PACO}/S_{\rm AT20G})$. The distribution of the latter quantity is better fitted by a Gaussian than that of the former one, see Fig~\ref{fig:gau}. The dotted line is the Gaussian best-fit, centered at -0.03 with $\hbox{FWHM}=0.34$, and the dashed lines are the mean and the $\hbox{mean}\pm \sigma$ of the distribution. The mean and the standard deviation, $\sigma$, of $\log_{10}(S_{\rm PACO}/S_{\rm AT20G})$ are $-0.043$ and $0.18$, respectively, corresponding to a dispersion $\sigma_v=\sigma(S_{\rm PACO}/S_{\rm AT20G})\simeq 0.40$, on a timescale of a few years. For comparison, for the bright sample \cite{MAS11} found, at 20 GHz, a variability of $\sim 9$ per cent at 20 GHz on a timescale of about 9 months, slightly larger than what has been found by \cite{SAD06} (6.9 per cent) over a one year timescale for a 100 mJy flux density limit sample. Our result is consistent with earlier indications (i.e., Ciaramella et al. 2004) that the variability increases with the time lag, for lags of up to several years.

\begin{table*}
\begin{center}
\caption{Fractions of PACO sources with $\simeq 19\,$GHz (more precisely, at 18.768 GHz) flux densities (mJy) within the bins indicated in the first column, found in the 32.768 (upper panel) and 39.768 GHz (lower panel) bins listed in the first row. The values in parenthesis are the number of sources in each 19 GHz bin.}
\label{tab:paco19paco33-40}
\begin{tabular}{|c|ccccccc}
\hline
\backslashbox{$S_{19\rm GHz}$}{$S_{33\rm GHz}$} & 200-250 & 250-400 & 400-750 & 750-1000 & 1000-1500 & 1500-2000 & 2000-5000 \\
\hline
200-250 (23)   & 0.11 & 0.   & 0.   & 0.   & 0.   & 0.   & 0.   \\
250-400 (38)   & 0.45 & 0.39 & 0.   & 0.   & 0.   & 0.   & 0.   \\
400-750 (68)   & 0.03 & 0.37 & 0.59 & 0.   & 0.   & 0.   & 0.   \\
750-1000 (28)  & 0.   & 0.   & 0.85 & 0.15 & 0.   & 0.   & 0.   \\
1000-1500 (35) & 0.   & 0.   & 0.17 & 0.45 & 0.38 & 0.   & 0.   \\
1500-2000 (15) & 0.   & 0.   & 0.   & 0.   & 0.60 & 0.40 & 0.   \\
2000-5000 (11) & 0.   & 0.   & 0.   & 0.   & 0.07 & 0.33 & 0.60 \\
\hline
\hline
\backslashbox{$S_{19\rm GHz}$}{$S_{40\rm GHz}$} & 200-250 & 250-400 & 400-750 & 750-1000 & 1000-1500 & 1500-2000 & 2000-5000 \\
\hline
200-250 (23)   & 0.11 & 0.   & 0.   & 0.   & 0.   & 0.   & 0.   \\
250-400 (38)   & 0.26 & 0.29 & 0.   & 0.   & 0.   & 0.   & 0.   \\
400-750 (68)   & 0.07 & 0.56 & 0.32 & 0.   & 0.   & 0.   & 0.   \\
750-1000 (28)  & 0.   & 0.12 & 0.85 & 0.03 & 0.   & 0.   & 0.   \\
1000-1500 (35) & 0.   & 0.   & 0.34 & 0.45 & 0.21 & 0.   & 0.   \\
1500-2000 (15) & 0.   & 0.   & 0.   & 0.   & 0.70 & 0.30 & 0.   \\
2000-5000 (11) & 0.   & 0.   & 0.   & 0.   & 0.13 & 0.47 & 0.40 \\
\hline
\end{tabular}
\end{center}
\end{table*}

An inspection of Fig.~\ref{fig:cfr} suggests that the dispersion $\sigma_v$ may be somewhat lower at the fainter flux densities. For the flux density interval $0.2\le S_{\rm AT20G}< 0.4\,$Jy we find $\langle\log_{10}(S_{\rm PACO}/S_{\rm AT20G})\rangle = -0.03$ and $\sigma[\log_{10}(S_{\rm PACO}/S_{\rm AT20G})]=0.17^{+0.01}_{-0.02}$, while for $S_{\rm AT20G}\ge 0.4\,$Jy we have $\langle\log_{10}(S_{\rm PACO}/S_{\rm AT20G})\rangle = -0.05$ and $\sigma[\log_{10}(S_{\rm PACO}/S_{\rm AT20G})] = 0.19^{+0.01}_{-0.01}$. The 68 per cent confidence errors on $\sigma$ have been computed following \cite{1980A&A....82..322D}. The difference is not statistically significant, as confirmed by a KS test that yields an 3.8 per cent probability that the two distributions come from the same parent population.

We have made the same analysis at lower frequency by comparing $S_{\rm PACO, 4.7GHz}$ with the flux densities at 4.8 GHz measured in the AT20G follow-up observations. The least square linear relation between the two sets of observations (see Fig.~\ref{fig:cfr_5}) gives $S_{\rm PACO}[Jy]= (0.92\pm 0.03) \cdot S_{\rm AT20G}[Jy]+(0.05\pm 0.04)$, confirming the pre- and post-CABB compatibility also at these frequencies. The variability ($\sigma_v$) is 38 per cent and the mean and standard deviation of the $\log_{10}(S_{\rm PACO}/S_{\rm AT20G})$ distribution are -0.006 and 0.15 respectively.

\subsection{Source Counts} \label{sec:count}
We use the simultaneous PACO measurements to estimate the distribution of flux densities at 33 or 40 GHz of sources with 19 GHz flux density within a given interval. To increase the statistics at $S_{\rm AT20G}\ge 500\,$mJy, we have taken into account also the PACO bright sample that covers an area a factor of 4 larger than the PACO faint sample. Using the results in Table~\ref{tab:paco19paco33-40}, we derive from the 20 GHz AT20G counts \citep{MAS10}, the counts at 33 and 40 GHz. The errors have been computed following \cite{GEH86}.

Table~\ref{tab:paco19paco33-40} suggests that additional, albeit small, contributions to the counts in the lowest and in the highest 33 or 40 GHz flux density bins could come from sources outside the flux density range covered by the PACO samples. Thus, strictly speaking, the derived counts in these bins are lower limits. While at low flux densities the incompleteness arises from the flux density limit of the PACO faint sample, at high flux densities it may be due to the limited sky coverage.

\begin{figure}
\begin{center}
\includegraphics[width=0.35 \textwidth, angle=90]{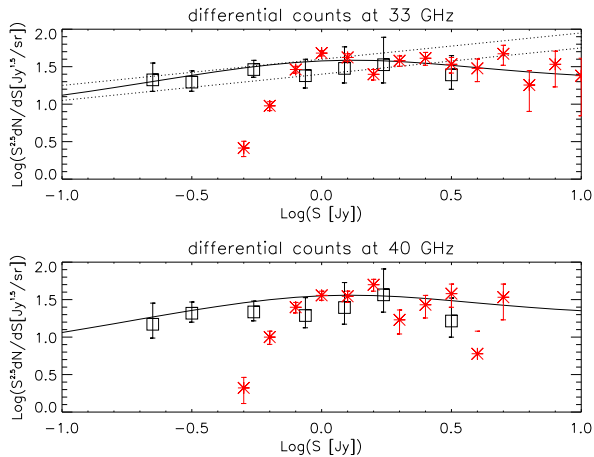}
\caption{Differential source counts at 33 and 40 GHz (squares) compared with the {\it Planck} counts \protect\citep{statprop} at 30 and 44 GHz (asterisks), and with the DASI estimate \protect\citep{KOV02} at 31 GHz (region bounded by dotted lines). The solid lines show the predictions of the \protect\cite{2005A&A...431..893D}.}
  \label{fig:counts}
\end{center}
\end{figure}

Figure~\ref{fig:counts} compares the present estimates of source counts (listed in Table~\ref{tab:counts}) with those obtained from the {\it Planck} ERCSC \citep{statprop} and from the DASI experiment \citep{KOV02}. The agreement is very good in the region of overlap, and our data extend the counts downwards in flux density by a substantial factor compared to those obtained by \cite{statprop}. The predictions of the \cite{2005A&A...431..893D} model are nicely consistent with these data.

\begin{table}
\caption{Source counts at 32.744 (second column) and 39.768 GHz(third column), with their errors, obtained by extrapolating the AT20G source counts.}
\centering
\label{tab:counts}
\begin{tabular}{ccc}
\hline
$\log(S)$ & $\log(S^{2.5}dN/dS)_{33\rm GHz}$ & $\log(S^{2.5}dN/dS)_{40\rm GHz}$ \\
$[\hbox{Jy}]$ & $[\hbox{Jy}^{1.5}/\hbox{sr}]$ & $[\hbox{Jy}^{1.5}/\hbox{sr}]$ \\
\hline
\hline
    -0.65 &       1.32 (+0.22, -0.15) &       1.17 (+0.28, -0.17) \\
    -0.50 &       1.29 (+0.15, -0.12) &       1.32 (+0.15, -0.12) \\
    -0.26 &       1.46 (+0.12, -0.10) &       1.33 (+0.15, -0.13) \\
    -0.06 &       1.38 (+0.22, -0.16) &       1.29 (+0.24, -0.16) \\
     0.09 &       1.48 (+0.29, -0.19) &       1.39 (+0.33, -0.22) \\
     0.24 &       1.53 (+0.37, -0.24) &       1.56 (+0.35, -0.23) \\
     0.50 &       1.39 (+0.26, -0.19) &       1.21 (+0.31, -0.22)   \\
\hline
\hline
\end{tabular}
\end{table}

The extension of the counts downwards in flux density is important to quantify the fluctuations in CMB maps due to sources below the detection limit, $S_l$. Since the differential counts below the completeness limits of the ERCSC ($\simeq 1\,$Jy at 30 GHz and $\simeq 1.5\,$Jy at 44 GHz, \cite{statprop}) go roughly as $dN/dS \propto S^{-2.2}$, the power spectrum of Poisson fluctuations due to sources fainter than $S_l$ goes as \citep{1996ApL&C..35..289D}:
\begin{equation}
C=\Omega \int^{S_l}_0 \frac{dN}{dS}\,  S^2\,  dS \propto S_l^{0.8},
\end{equation}
where  $\Omega$ is the beam solid angle. This means that $\ge 70$ per cent of the residual point source fluctuations in {\it Planck} 30 and 44 GHz maps, after subtraction of directly detected sources, are contributed by sources in the flux density range covered by the PACO survey. Note that fluctuations due to clustering, not taken into account in the above equation, are negligible in the case of radio sources because of the strong dilution due to the broadness of the luminosity function.

\begin{figure}
\begin{center}
\includegraphics[width=0.35 \textwidth, angle=90]{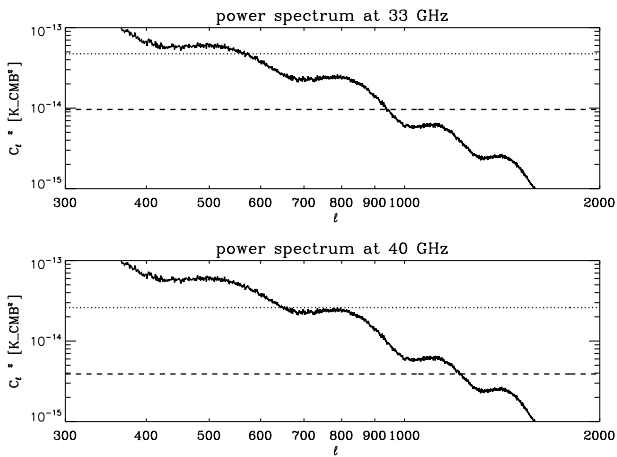}
\caption{Contribution of the extragalactic point sources to the CMB power spectrum (black solid line) as a function of multipole (l) for different flux density limits. In each panel the upper (dot-dashed) line refer to the ERCSC completeness limits (1 Jy at 33 GHz and 1.5 Jy 40 GHz), while the lower (dashed line) refers to the flux density limit of the PACO faint sample (200 mJy).}
  \label{fig:CL}
\end{center}
\end{figure}

The effect of an assessment of the counts down to fainter flux limits on the contribution to the power spectrum of radio sources below the detection limit is shown in Fig.~\ref{fig:CL}. Here the dotted line refers to the completeness limits of the ERCSC \citep{ERCSC}, 1 Jy at 33 GHz and 1.5 Jy at 40 GHz \citep{statprop}. The present assessment of the counts down to 200 mJy decreases the amplitude of the extragalactic source power spectrum as shown by the dashed lines.

\section{Conclusions} \label{sec:conclusion}
The PACO faint sample is a complete sample comprising 159 AT20G sources with $S_{20\rm GHz}>200\,$mJy. The sources have been observed between July 2009 and August 2010 in 3 frequency bands of $2\times 2$ GHz each, centered at 5.5, 9, 18, 24, 33, 39 GHz. Most observations were carried out within 10 days from {\it Planck} observations. The {\it Planck} ERCSC \citep{ERCSC} reports the detection at least at one frequency of 57 PACO faint sources. Of these, 25 have measurements at more than 5 {\it Planck} frequencies. Some examples of spectra obtained combining PACO and {\it Planck} observations are shown in Fig.~\ref{fig:allsed}. There are 4 ERCSC sources within the PACO faint area not included in our sample. One is Fornax A, missed by the AT20G because its emission extends over an area larger than the ATCA field-of-view. The other 3 are at low Galactic latitude and were not detected by the AT20G survey. They may be either extended Galactic sources or spurious (peaks of the Galactic diffuse emission).  Thus the fraction of potentially spurious ERCSC sources in the PACO faint area is $\le 5\%$.
Many more sources should be detected by {\it Planck} as new surveys are completed and the data analysis improves. Additional constraints on the source spectra at {\it Planck} frequencies will be extracted from {\it Planck} maps, when they will become available, by a stacking analysis.

A comparison with AT20G measurements, carried out, on average, a few years earlier, has demonstrated that, on these timescales, our sources show a rather high variability with an rms amplitude of $\simeq 40$ per cent at 20 GHz.

The analysis of the spectra, within the PACO frequency range, of the 98 sources in the PACO faint sample with $S_{20\rm GHz}<500\,$mJy (i.e. below the flux density limit of the bright PACO sample) shows no significant differences with the bright sample \citep{MAS11}. The main difference is an increase of the percentage of steep sources from 3.6 per cent in the bright sample to 13.3 per cent in the faint one. The high-frequency steepening found by \cite{MAS08} and by \cite{MAS11} is confirmed.

Our data have allowed us to extrapolate the 20 GHz source counts to 33 and 40 GHz, down to $\simeq 200\,$mJy, i.e. a factor $\ge 5$ below current estimates from {\it Planck} data at nearby frequencies (30 and 44 GHz). Our counts are in very good agreement with the {\it Planck} ones in the region of overlap and are well accounted for by the \cite{2005A&A...431..893D} model. The assessment of the counts down to faint flux density limits substantially improves the control of fluctuations due to unresolved sources in {\it Planck} maps.

\section*{Acknowledgments}
We are grateful to the editor and to the referee for a careful reading of the manuscript and for helpful comments.

MM, AB, JGN and GDZ acknowledge financial support for this research by ASI (ASI/INAF Agreement I/072/09/0 for the {\it Planck} LFI activity of Phase E2 and contract I/016/07/0 COFIS). RDE acknowledges support of a Federation Fellowship (FF0345330).

We thank the staff at the Australia Telescope Compact Array site, Narrabri (NSW), for the valuable support they provide in running the telescope. The Australia Telescope Compact Array is part of the Australia Telescope which is funded by the Commonwealth of Australia for operation as a National Facility managed by CSIRO.


\end{document}